\documentstyle[aps,multicol,prb,epsf]{revtex} 
 
\begin{document} 
\draft 
 
\title{Magnon-assisted Andreev reflection in
a ferromagnet-superconductor junction} 
 
\author{Edward McCann and Vladimir I. Fal'ko} 

\address{Department of Physics, Lancaster University,
Lancaster, LA1 4YB, United Kingdom}

\date{\today} 

\maketitle \begin{abstract} {
We study subgap transport in a ferromagnet-superconductor
junction at low temperature due to Andreev reflection.
The mismatch of spin polarized current in the ferromagnet and
spinless current in the superconductor results in an additional
contact resistance which can be reduced by magnon emission
in the ferromagnet. 
Using the s-f model and focusing on half-metallic ferromagnets,
we calculate the corresponding nonlinear contribution to the
current which is asymmetric with respect to the sign of the
bias voltage and is related to the density of states of magnons
in the ferromagnet.
}\end{abstract}

 
\begin{multicols}{2}
\bibliographystyle{simpl1}

Spin polarized transport is
a subject of intense research,
motivated largely by the possibility
of developing a form of electronics which utilizes the
spin polarization of carriers\cite{pri95}.
Ferromagnetic (F) metals have more carriers of one spin polarization
(known as majority carriers) present at the Fermi energy $E_F$
than of the inverse polarization (minority carriers).
Of particular interest are so called half-metals where the
splitting $\Gamma$ between the majority and minority conduction bands
is greater than $E_F$ measured from the bottom of the majority band.
In such a material only majority carriers are present at the
Fermi energy and electric current in it is completely spin polarized.
An F-F junction with antiparallel spin polarizations
has a larger contact resistance than a junction with parallel
polarizations due to a mismatch of spin currents at
the interface, leading to tunneling magnetoresistance
in F-F junctions\cite{jul75,moo00}
and giant magnetoresistance in multilayer structures\cite{bai88,pra91}.
Spin current mismatch may also affect the conductance of a
ferromagnet-superconductor junction\cite{deJ95,Fal99a,Jed99,Fal99b}.
At low temperatures and small bias voltage, current flows
through the interface due to Andreev reflection\cite{And64}
whereby particles in the ferromagnetic region with excitation energies
$\epsilon$ smaller than the superconducting gap energy $\Delta$
are reflected from the interface as holes.
Since subgap transport in the superconductor (S) is mediated by
spinless Cooper pairs the spin current is zero in the
superconductor in contrast to the ferromagnet.

Spin relaxation processes\cite{pri95,j+s85},
such as spin-orbit scattering at impurities
or magnon emission, can reduce the spin current mismatch.
In a given junction, spin-orbit scattering
(which is an elastic process) would reduce the value of the additional
contact resistance \cite{Fal99b,Bax99} whereas the inelastic process of
magnon emission would manifest itself as a modification of the
form of the $I(V)$ characteristics.
Nonlinear $I(V)$ characteristics due to
magnon-assisted tunneling between two ferromagnets
have already been studied both theoretically\cite{bra98}
and experimentally\cite{tsu71}
with a view to relate the second derivative of the current
to the density of states of magnons $\Omega (\omega)$
as $d^2I/dV^2 \propto \Omega (eV)$.
In the present work we analyze the role of magnon
emission in the formation of the subgap $I(V)$ characteristics, $eV < \Delta$,
of a ferromagnet-superconductor junction with emphasis
given to half-metallic ferromagnets.
Microscopically, the relevant process involves the transfer
of a singlet pair of electrons from the superconductor to the
ferromagnet where one of them forms an intermediate state at an energy
below the bottom of the conduction band for minority electrons.
Then, this latter electron can emit a magnon in the ferromagnet
which will carry away excess spin allowing the electron to incorporate
itself into the majority conduction band.
The s-f model\cite{S+M81,W+W70} is employed
to describe a ferromagnet assuming that
electrical conduction and magnetism are caused by different
groups of electrons which interact via s-f exchange.

To anticipate a little,
due to magnon-assisted Andreev reflection at zero temperature,
the $I(V)$ characteristics of an F-S junction for $eV < \Delta$
acquire an additional contribution which is specified by the
differential of the nonlinear differential conductance,
\begin{eqnarray}
\frac{d^2I}{dV^2} &=&
\frac{e  v  G_{A}}{S}
\left\{ \frac{\Omega (2eV)}
{\left[
1 + \frac{(2eV - \omega_0)W}{\Gamma\omega_m}
\right]^2
\left[ 1 - \left( \frac{eV}{\Delta} \right)^2 \right]} \, + \right.
\nonumber \\
&&
+ \left.
\int_{0}^{2eV}
\frac{\Omega (\epsilon ) 
\left[
F \left( \frac{eV}{\Delta} \right)
-
F \left( \frac{\epsilon - eV}{\Delta} \right)
\right] \frac{d\epsilon}{2\Delta}}
{\left[
1 + \frac{(\epsilon - \omega_0)W}{\Gamma\omega_m}
\right]^2}
\right\} .
\label{d2i}
\end{eqnarray}
Here $F(x) = x/(1-x^2)^2$,
$\omega_m$ is the magnon bandwidth
and $\omega_0$ is the anisotropy energy responsible for a gap in the
magnon density of states [$\Omega (\omega )$ is non-zero for
$\omega_0 < \omega < \omega_m$ and zero otherwise]\cite{note:sing}.
The factor of two in $\Omega (2eV)$ arises because
an elementary Andreev reflection process
involves a net transfer of charge $2e$ across the biased junction.
The conductance $G_{A}$ incorporates tunneling parameters
relevant for the Andreev process and it is defined explicitly later.
$W$ is the conduction band width of electrons,
$v$ is the volume of a unit cell, and
$\Gamma = 2{\cal J}S$ is the conduction electron band splitting
where $S$ is the spin of the local moments,
and ${\cal J}$ is the s-f exchange constant.

The nonlinear contribution to the current, Eq.~(\ref{d2i}),
is asymmetric with respect to bias voltage $V$.
It is zero for negative bias and finite
for $V > \omega_0 /(2e)$.
This feature can be expained using the sketch in Fig~1
which illustrates the tunneling process 
between an S electrode on the left hand side
and an F electrode on the right for
(a) $V>0$ and (b) $V<0$.
We have adopted the convention
that positive (negative) voltage results in a Fermi energy $E_F$ in
the ferromagnet that is lower (higher) by energy $|eV|$ than the
Fermi energy $E_S$ in the superconductor.
For $V>0$, Fig~1(a), Andreev reflection
results in the injection of both a majority (spin `up') and a minority
(spin `down') electron into the ferromagnet.
One of these electrons has an energy $\epsilon$ above the Fermi energy
in the superconductor $E_S$,
the other has an energy $\epsilon$ below $E_S$.
With respect to the Fermi energy in the ferromagnet $E_F$ these
energies are $eV + \epsilon$ and $eV - \epsilon$, respectively, which are both
above $E_F$ because of the need to move into unoccupied states in
the ferromagnet, $\epsilon \leq eV$.
At zero temperature $T=0$ the core spins are all aligned in the majority,
up direction and, if $E_F \ll \Gamma$, then only
spin up conduction electrons are present at $E_F$ in the ferromagnet.
A dynamic process which allows a spin down electron to enter the ferromagnet
is shown schematically in Fig~1(a).
The spin down electron tunnels from the superconductor into
a virtual, intermediate spin down state above $E_F$, then it emits a magnon
and incorporates itself into an empty state in the majority conduction band.
In Fig~1(a) the magnon is depicted as a flip
in the spin of one of the localized magnetic moments.

On the other hand, for $V<0$, Fig~1(b), Andreev reflection
would result in the injection of a spin up and a spin down electron
from the ferromagnet into the superconductor
with energies $\epsilon$ above and below $E_S$.
With respect to the Fermi energy in the ferromagnet $E_F$ these
energies are $-|eV| + \epsilon$ and $-|eV| - \epsilon$,
respectively, which are both
below $E_F$ because of the need to have initially occupied states in
the ferromagnet, $\epsilon \leq |eV|$.
As stated above, there are no spin down states at these energies near $E_F$.
Since a spin up electron cannot emit a magnon (left side of Fig~1(b)),
due to conservation of total spin in the exchange interaction,
the only possibility would be that a spin up electron in the ferromagnet
would absorb a magnon to form an intermediate spin down state
before tunneling into the superconductor.
However there are no thermally excited magnons at $T=0$
in the initial state of the ferromagnet (right side of Fig~1(b))
so it is impossible for magnon-assisted Andreev reflection to
contribute to current formation in negatively biased junctions.

To calculate the current across an F-S junction, we use the
transfer Hamiltonian method and the nonequilibrium Keldysh
technique\cite{car72,cue96} to express the current in terms
of the tunneling density of states in the metallic ferromagnet
which will be determined using the s-f model.
We will generalize the approach of Ref.~\onlinecite{cue96}
developed for normal-superconductor junctions
to the case of an F-S junction.
We consider a point contact geometry which consists of a narrow
constriction between S and F electrodes and we use
the total Hamiltonian of the system in the form of
\begin{eqnarray}
H &=& H_S + H_F + H_T   ,  \nonumber \\
H_T &=& \sum_{\alpha}
\left(
t \, c_{S,\alpha}^{\dag} c_{F,\alpha} + 
t^{*} \, c_{F,\alpha}^{\dag} c_{S,\alpha}
\right) , \nonumber
\end{eqnarray}
in all calculations paying attention to the
spin index $\alpha = \{ \uparrow ,\downarrow \}$.
In the tunneling Hamiltonian $H_T$, $t$ describes the transfer
of an electron from S to F.
We assume that $t$ is independent of energy and that spin is
conserved upon electron transfer across the interface.

Assuming that the voltage drop $V$ occurs across the point contact
and both electrodes are almost in equilibrium, for a given
distribution and density of states in the reservoirs,
the current can be calculated as the mean value of the current operator,
$I = (ie /\hbar ) \sum_{\alpha}
(t \, \langle c_{S,\alpha}^{\dag} c_{F,\alpha} \rangle - 
t^{*} \, \langle c_{F,\alpha}^{\dag} c_{S,\alpha} \rangle )$.
The averages in this equation
may be expressed in terms of
nonequilibrium Keldysh Greens functions $\hat{G}^{+-}$
and it is convenient for the case of a superconducting electrode
to use the $2 \times 2$ Nambu representation\cite{cue96}.
After some algebra we arrive at the following expression for
the current
\begin{eqnarray}
I (V) &=& \frac{e}{h}
4\pi^2 \left| t \right|^4
\int_{-\infty}^{\infty} d\epsilon
\left| \hat{G}_{S,12}^{R} (\epsilon ) \right|^2  \nonumber \\
&& \times \, 
\left[ f (\epsilon - eV) - f (\epsilon + eV) \right] \nonumber \\
&& \times \, \left[
\rho_{\uparrow} (E_F+ \epsilon + eV) \rho_{\downarrow} (E_F - \epsilon + eV)
+ \right. \nonumber \\
&& \qquad \left. + \,
\rho_{\downarrow} (E_F+ \epsilon + eV) \rho_{\uparrow} (E_F - \epsilon + eV)
\right] .
\label{tunfs}
\end{eqnarray}
This expression describes the current due to Andreev reflection
of electrons at the F-S interface.
The gap-full spectrum of quasi-particles in the superconductor
is taken into account in the off-diagonal element of the retarded
Greens function $\hat{G}_{S,12}^{R}$ and the energy $\epsilon$ is 
measured with respect to the Fermi energy $E_S$ in the superconductor.

In Eq.~(\ref{tunfs}), $\rho_{\alpha} (E)$ is the tunneling density of states
of spin state $\alpha$ in the ferromagnet.
For $\Gamma \gg \{ E_F,eV\}$ and in the absence of spin waves,
$\rho_{\downarrow} (E)$ is zero in a half metal at energies close to the
chemical potentials of the electrodes and $I(V) = 0$.
However, as explained above, $\rho_{\downarrow} (E)$ is nonzero for
$E > E_F + \omega_0$ because of the dynamical process
of magnon-assisted tunneling.
The tunneling density of states $\rho_{\downarrow} (E)$ is determined
using the s-f model\cite{S+M81,W+W70} with the following Hamiltonian
\begin{eqnarray}
H_F &=& H_0 + H_J + H_{ex} , \nonumber \\
H_0 &=& \sum_{\langle ij \rangle\alpha} t_{ij} c_{i\alpha}^{\dag} c_{j\alpha}
\equiv \sum_{k\alpha} \epsilon_k c_{k\alpha}^{\dag} c_{k\alpha} ,\nonumber \\
H_J &=& - J \sum_{\langle ij\rangle} {\bf S_{i}.S_{j}}
- \omega_0 \sum_{i} S_{i}^{z} ,\nonumber \\
H_{ex} &=& - {\cal J} \sum_{j\alpha\beta}
\left({\bf \sigma_{\alpha\beta}.S_{j}}\right)
c_{j\alpha}^{\dag} c_{j,\beta} ,  \nonumber 
\end{eqnarray}
where $H_0$ deals with conduction band electrons,
$H_J$ is the Heisenberg Hamiltonian of the localized moments
and $H_{ex}$ takes into account the s-f intra-atomic exchange.
We will consider weak, ferromagnetic s-f coupling $0 < {\cal J}S < W$,
thus excluding the case corresponding to double exchange ${\cal J}S \gg W$.

The conduction electron single particle Greens function and energy
spectrum have been studied in the limit of zero temperature both
for zero conduction band filling\cite{S+M81}
and finite filling\cite{W+W70}.
In the regime $\left\{eV,\omega_0\right\} \ll E_{F} \ll \Gamma \ll W$
the minority electron density of states is
\begin{equation}
\rho_{\downarrow} (E)
\approx
\frac{v\Gamma^2}{2S L^d}
\sum_{k^{\prime}} \sum_q
\frac{\left[ 1 - f(k^{\prime}) \right]
\delta \left( 
\epsilon_{k^{\prime}} - E + \omega_q
\right)}
{\left(
\epsilon_{k^{\prime}+q} - E + \Gamma
\right)^2}   ,
\label{downdos0}
\end{equation}
where the energy $E$ is measured from the bottom of the spin up
conduction band and $L^d$ is the volume of the system.
A similar expression, with $2S = 1$, was also obtained
for itinerant ferromagnets using the Hubbard model\cite{E+H73}.
The energy in the denominator of Eq.~(\ref{downdos0}),
$\epsilon_{k^{\prime}+q} - E + \Gamma$,
is related to the inverse lifetime of an electron in the virtual state,
the delta function in the numerator accounts for energy conservation
of the entire process
and the factor $[1 - f(k^{\prime})]$ indicates that this process
is a dissipative one and only involves unoccupied final (spin up) states.
Therefore the minority density of states is finite for energies
$E > E_F + \omega_0$ only.
At $T=0$, in our regime of interest,
spin waves do not affect the majority density of states
$\rho_{\uparrow} (E) \approx N (E)$,
where $N (E)$ is the bare electron density of states arising from
the Hamiltonian $H_0$.
We have neglected interactions between conduction electrons
so that the dominant effect
of finite filling is the requirement to obey Fermi statistics.

The dispersion of magnons in a ferromagnet is quadratic at low
frequencies, $\omega_q = Dq^2 + \omega_0$ where
$D = 2JS^2a^2$ and $a$ is the lattice constant.
For simplicity, we shall also use the Debye approximation
for a broader range of energies of the magnon spectrum
by introducing a cutoff of the parabolic magnon dispersion
at $\omega_m \approx D (6\pi^2 /v)^{2/3}$
since, for $\Delta \ll \omega_m$, such a simplification
does not influence the final results.
We also use a parabolic approximation for free electrons in
a conduction band and find that,
for a three dimensional ferromagnetic metal and $E \approx E_F$,
\begin{eqnarray}
\rho_{\downarrow} (E)
&\approx&
\frac{v N (E)}{2S}
\int_{0}^{E-E_F}
\frac{\Omega (\omega ) \, d\omega}
{\left[ 1 + \left(\frac{W}{\Gamma}\right)
\frac{\omega - \omega_0}{\omega_m} \right]^{2}} ,
\label{downdos}  \\
\Omega (\omega ) &=&
\frac{\sqrt{\omega -\omega_0}}{4\pi^2D^{3/2}} \quad ;
\qquad \omega_0 < \omega < \omega_m .
\end{eqnarray}

We use Eq.~(\ref{tunfs}) to calculate the current due to
Andreev reflection for low bias voltage $eV < \Delta$,
taking into account the lowest order
in the tunneling matrix elements, $|t|^4$.
In a BCS superconductor
the off-diagonal Greens function is
$\hat{G}_{S,12}^{R} (\epsilon )\approx \pi N(E_S)
/\sqrt{ 1 - (\epsilon /\Delta )^2}$
for $\epsilon < \Delta$,
where $N(E_S)$ is the density of states in the normal state
near the Fermi energy.
Note that if the ferromagnet were to be replaced by a normal metal
where the density of states at $E_F$ is $N (E_{F} )$ for
both spin channels, the current at low bias would be $I = VG_{A}$ with
conductance
$G_{A} \approx 
(e^2/h)
16 \pi^4 \left| t \right|^4
N^2 (E_{F} ) N^2 (E_{S} )$.
As we assume $eV \ll E_F$, $\rho_{\uparrow} \approx N(E_F)$,
and Eq.~(\ref{tunfs}) for an F-S junction can be simplified,
by shifting the energy variable, into
\begin{equation}
I (V) \approx
\frac{G_{A}}{2e N(E_F)}
\int_{0}^{2eV}
\frac{d\epsilon \, \rho_{\downarrow} (E_F + \epsilon )}
{\left[ 1 - \left( \frac{eV - \epsilon}{\Delta}\right)^2\right]} .
\label{i1}
\end{equation}
Using $\rho_{\downarrow} (E_F + \epsilon )$ from
Eq.~(\ref{downdos}), we find for
$eV < \Delta \ll \omega_m \Gamma /W$ that\cite{note:sing}
\begin{eqnarray}
I (V) &\approx&
\frac{v \, \Delta \, G_{A}}{24 e \pi^2 S D^{3/2}}
\theta (2eV-\omega_0)
\Bigg\{ \!\!
- 2\Delta \sqrt{2eV - \omega_0} +
\nonumber \\
&& \!\!\! \!\!\!
+
\left( \Delta - eV + \omega_0 \right)^{3/2}
\arctan \left( \sqrt{\frac{2eV - \omega_0}{\Delta - eV + \omega_0}} \right)
\nonumber \\
&& \!\!\! \!\!\! 
+
\left( \Delta + eV - \omega_0 \right)^{3/2}
\mathrm{arctanh}
\left( \sqrt{\frac{2eV - \omega_0}{\Delta + eV - \omega_0}} \right)
\Bigg\} . \nonumber
\end{eqnarray}
The $I(V)$ characteristics obtained yield the relation between the
magnon density of states and $d^2 I/dV^2$ shown in Eq.~(\ref{d2i}).
For small voltages $eV \ll \Delta$, the current may be
simplified further as
\begin{equation}
I (V) \approx
\frac{v \, G_{A}}{60 e \pi^2 S D^{3/2}}
(2eV-\omega_0)^{5/2} \, \theta (2eV-\omega_0) ,
\end{equation}
and $d^2 I/dV^2 \propto \Omega (2eV)$.
The factor $\theta (2eV-\omega_0)$ emphasises the absence of a
contribution to the current from magnon-assisted Andreev
reflection for negative bias.
At zero temperature, the $I(V)$ characteristics are asymmetric,
which can be used to extract magnon-assisted Andreev reflection
in an F-S junction involving a ferromagnet with electrons of
both polarizations present at $E_F$.
Note that at $T > \omega_0$, the absorption of thermal magnons
would allow for a similar contribution to the current at $V < 0$.
A graphic representation of the result of Eq.~(\ref{d2i}) is
given in Fig~2, in comparison with a similar analysis
for an F-F junction in an antiparallel configuration (inset).
In particular, we illustrate how the relation between the
conduction electron bandwidth and the exchange splitting energy,
$W /\Gamma$, may affect the result, since
for $W \gg \Gamma$ the contribution of large energy magnons
to the lifting of spin current mismatch is not so
efficient as for $W \ll \Gamma$.

To summarise, spin current mismatch in an F-S junction at subgap voltages
may be lifted by magnon-assisted tunneling resulting in
nonlinear $I(V)$ characteristics related to the magnon density of states.
Such features are typical not only for junctions between ferromagnets
and BCS superconductors, but also for HiTc materials, since
subgap transport through any spin singlet superconductor
would be affected by spin current mismatch at the F-S interface.
For a matching pair of materials, with the Curie temperature
of the ferromagnet similar to the transition temperature
of the HiTc material,
the Andreev reflection process could be used to probe the whole magnon
spectrum.

%

The authors thank N.~R.~Cooper, D.~M.~Edwards, C.~J.~Lambert,
and Yu.~V.~Nazarov for discussions.
This work was supported by EPSRC, with partial travel support
from NATO and COST.


%
\begin{figure}
\hspace{0.08\hsize}
\epsfxsize=0.78\hsize
\epsffile{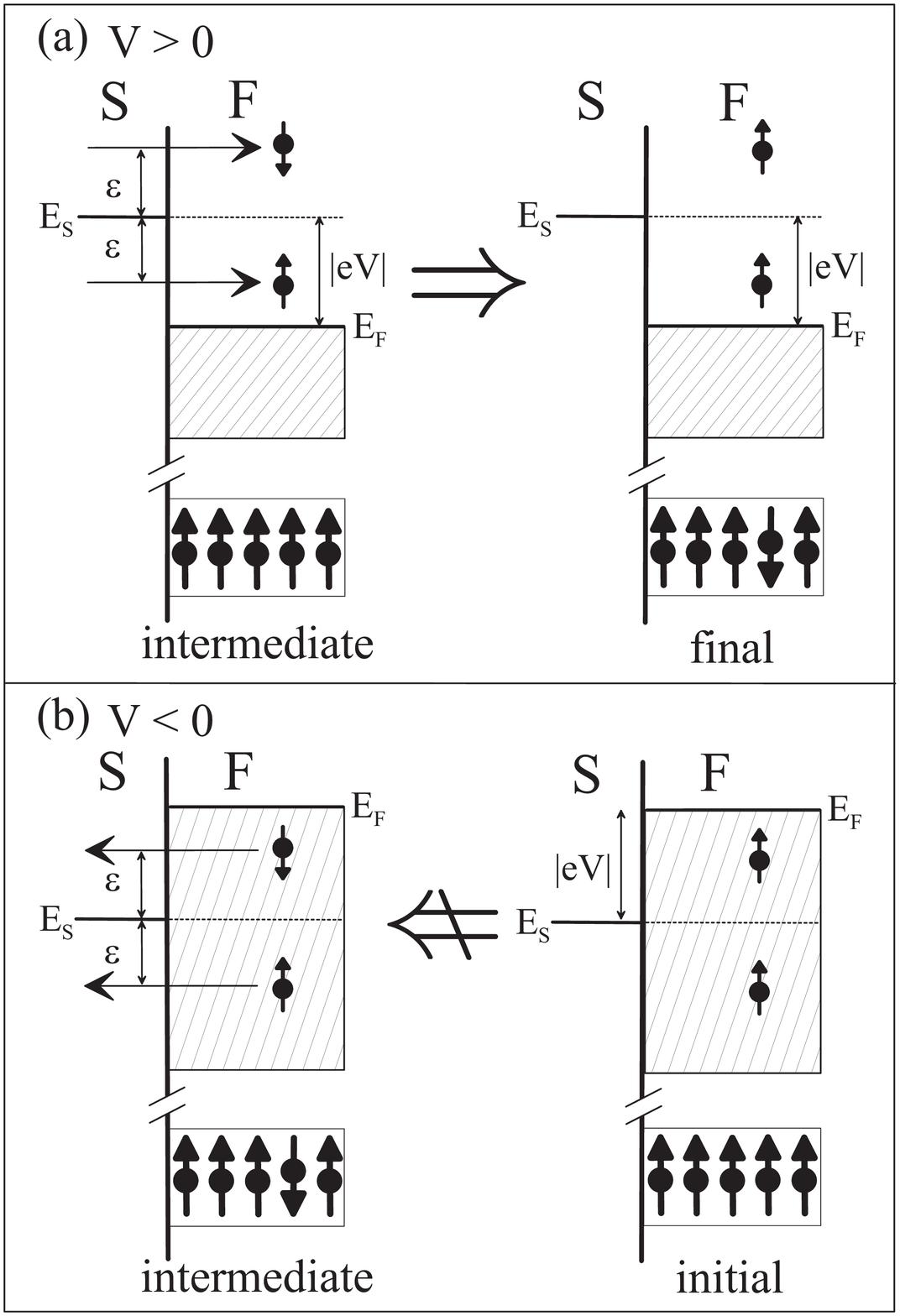}
\refstepcounter{figure}
\label{figure:1}

{\setlength{\baselineskip}{10pt} FIG.\ 1.
Schematic of the tunneling process 
between a superconducting electrode on the left hand side
and a ferromagnetic electrode on the right for
(a) $V>0$ and (b) $V<0$.
For (a) $V>0$ a down spin electron tunneling into the ferromagnet
may emit a magnon and incorporate itself into the majority conduction band.
For (b)  $V<0$ no spin flip process is possible at $T = 0$ because
in the initial state (right) there
are no thermal magnons for an up spin electron to absorb.
See text for details.}
\end{figure}
%
%
\begin{figure}
\epsfxsize=1.0\hsize
\epsffile{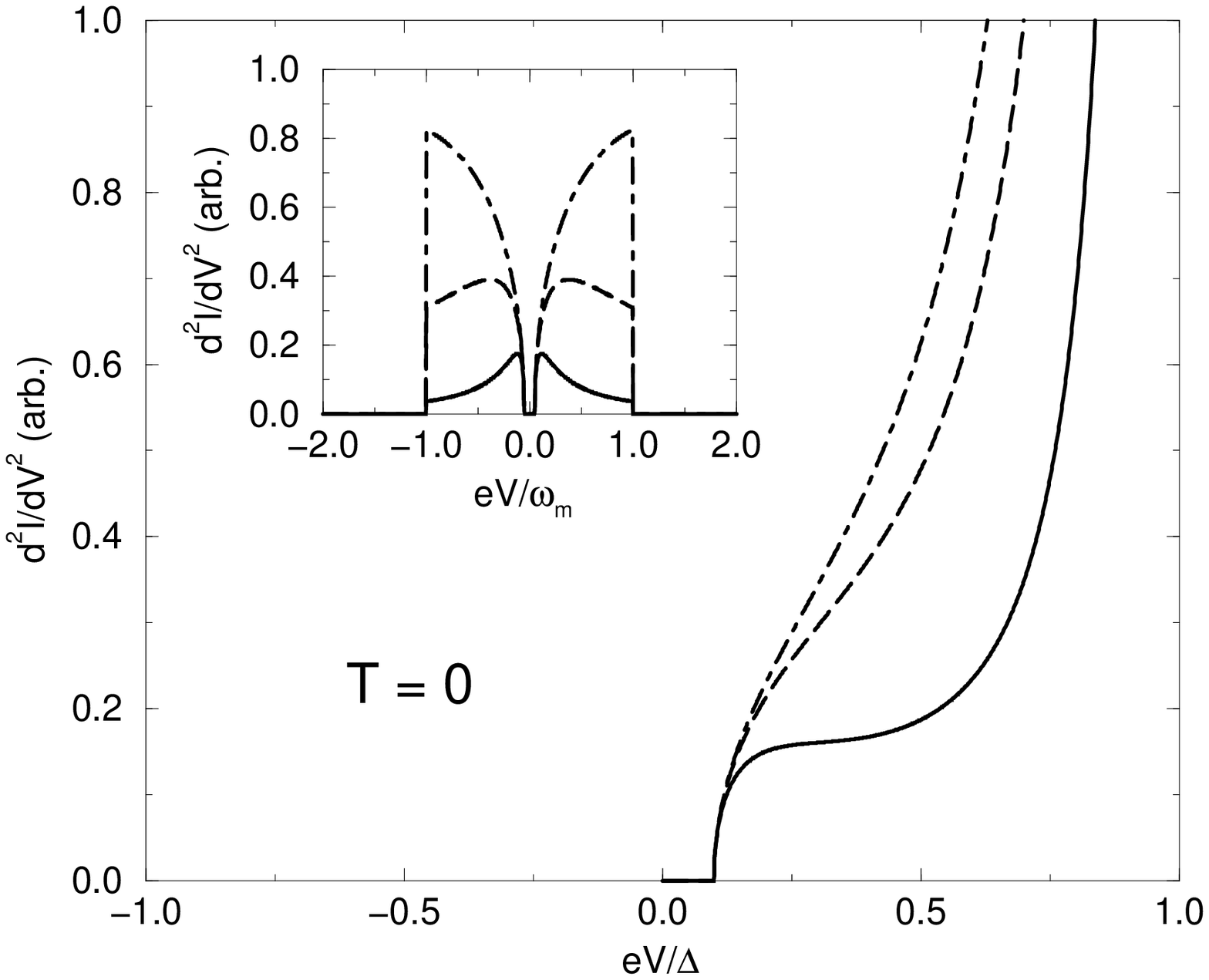}
\refstepcounter{figure}
\label{figure:2}
{\setlength{\baselineskip}{10pt} FIG.\ 2.
Typical form of $d^2I/dV^2$ for an F-S junction with various choices
of material parameters.
From top to bottom, $\Gamma/W = 5.0$ (dot-dashed), $1.0$ (dashed),
and $0.2$ (solid).
As an example, we choose $\omega_0 = 0.2 \Delta$
and $\omega_m = 4.0 \Delta$.
For comparison, inset shows $d^2I/dV^2$ for an F-F junction with antiparallel
spin polarizations.
}
\end{figure}
%

\end{multicols}


\begin{references}
\bibitem{note:sing}
Strictly speaking, the condition $eV < \Delta$ should be replaced by
$eV < \Delta \sqrt{1 - \pi^2|t|^2 N^2(E_F)}$
where we consider $\pi^2|t|^2 N^2(E_F) \ll 1$.
\end{references}
\end{document}